\documentclass[journal,transmag]{IEEEtran}

\usepackage{graphicx}
\usepackage{amsmath}
\usepackage{amssymb}
\usepackage{caption2}
\usepackage{booktabs}
\usepackage{multirow}
\usepackage{makecell}
\usepackage{adjustbox}
\usepackage{enumitem}
\usepackage{hyperref}
\usepackage{marvosym}
\usepackage{algorithm}
\usepackage{amssymb}
\usepackage{algorithmicx}
\usepackage{algpseudocode}
\usepackage{diagbox}
\usepackage{graphicx,bm,subfigure}
\usepackage{cite}
\usepackage{braket}
\usepackage{tikz}
\usetikzlibrary{calc}

\ifCLASSINFOpdf

\else

\fi

\floatname{algorithm}{Algorithm}

\begin{document}

\title{DCF-DS: Deep Cascade Fusion of Diarization and Separation for Speech Recognition under Realistic Single-Channel Conditions}

\author{Shu-Tong Niu, Jun Du,~\IEEEmembership{Senior Member,~IEEE}, Ruo-Yu Wang, Gao-Bin Yang, Tian Gao, Jia Pan and Yu Hu}

\markboth{}
{Shell \MakeLowercase{\textit{et al.}}: Bare Demo of IEEEtran.cls for IEEE Transactions on Magnetics Journals}

\maketitle

\begin{abstract}
We propose a single-channel Deep Cascade Fusion of Diarization and Separation (DCF-DS) framework for back-end automatic speech recognition (ASR), combining neural speaker diarization (NSD) and speech separation (SS). First, we sequentially integrate the NSD and SS modules within a joint training framework, enabling the separation module to leverage speaker time boundaries from the diarization module effectively. Then, to complement DCF-DS training, we introduce a window-level decoding scheme that allows the DCF-DS framework to handle the sparse data convergence instability (SDCI) problem. We also explore using an NSD system trained on real datasets to provide more accurate speaker boundaries. Additionally, we incorporate an optional multi-input multi-output speech enhancement module (MIMO-SE) within the DCF-DS framework, which offers further performance gains. Finally, we enhance diarization results by re-clustering DCF-DS outputs, improving ASR accuracy. By incorporating the DCF-DS method, we achieved first place in the realistic single-channel track of the CHiME-8 NOTSOFAR-1 challenge. We also perform the evaluation on the open LibriCSS dataset, achieving a new state-of-the-art single-channel speech recognition performance.
\end{abstract}


\IEEEdisplaynontitleabstractindextext

\IEEEpeerreviewmaketitle

\begin{tikzpicture}[remember picture, overlay]\node at ($(current page.north) + (0,-0.5in)$)[text width=18cm] {\small This work has been submitted to the IEEE for possible publication.\\ Copyright may be transferred without notice, after which this version may no longer be accessible.};
\end{tikzpicture}

\section{Introduction}
\label{section1: introduction}
Multi-speaker speech recognition aims to identify ``who spoke what and when" in recordings with unknown scenarios and speakers \cite{raj2021integration}. While deep learning has driven significant advancements in speech recognition \cite{prabhavalkar2023end, deng2013new}, accurate speech recognition in real-world multi-speaker conversational scenarios remains challenging. The difficulties often arise from high overlap rates, background noises and reverberation, unknown speaker numbers, and natural conversation styles \cite{boeddeker2024ts, cornell23_chime, vinnikov24_interspeech}. In this context, speaker diarization and speech separation are commonly employed as pre-processing steps. Specifically, speaker diarization segments the audios based on speaker identity \cite{2021review, review2}, and speech separation \cite{separation1} isolates each speaker’s speech from the mixed audio signals.


Speaker diarization and speech separation are closely related tasks \cite{boeddeker2024ts}, as they both aim to predict the distribution of speakers: speaker diarization predicts the speaker's presence across time frames, whereas speech separation predicts their distribution across time-frequency (T-F) bins. Consequently, the temporal distribution provided by diarization can serve as a prior for speech separation, helping the system to better distinguish between speakers \cite{boeddeker2018front, medennikov2020stc, niuISSD, niu2023qdm, boeddeker2024ts, taherian2024multi}. Speech separation, in turn, inherently encompasses diarization—ideal separation results reveal the time regions of speakers \cite{ fang2021deep, von2019all, kinoshita2020tackling}. Moreover, speech separation can handle overlapping segments, which can enhance speaker diarization accuracy \cite{chen2020continuous, vinnikov24_interspeech}. 
Due to the strong correlation between speaker diarization and speech separation, numerous kinds of research \cite{lin2021sparsely, maiti2023eend, kalda24_odyssey, chen2020continuous, vinnikov24_interspeech, fang2021deep, von2019all, kinoshita2020tackling, boeddeker2018front, medennikov2020stc, niuISSD, niu2023qdm, boeddeker2024ts, taherian2024multi, ao2023used} have been conducted to explore combining these two tasks to enhance speech quality. Based on how the two tasks are combined, these methods can be broadly categorized into multi-task joint training and sequential cascade approaches. 

In multi-task joint training approaches \cite{lin2021sparsely, maiti2023eend, ao2023used, kalda24_odyssey}, the network typically generates two outputs: speaker activity probabilities and separated speech. This structure allows the two tasks to benefit from each other mutually \cite{maiti2023eend}. Speaker diarization results can also be used to post-process the separation outputs, reducing background noises in the separated results \cite{lin2021sparsely, maiti2023eend, kalda24_odyssey, ao2023used}. Among these methods, some studies \cite{lin2021sparsely, ao2023used} propose a network that jointly performs target speech extraction and voice activity detection (VAD), yielding improvements in both fully and sparsely overlapped speech. Another approach, end-to-end neural speaker diarization and separation (EEND-SS) \cite{maiti2023eend}, integrates speaker counting, diarization, and separation in a multi-tasking framework. 
However, due to the inclusion of the separation task within the joint training framework, these methods are restricted to training on simulated data. Unsupervised training methods, such as PixIT \cite{kalda24_odyssey}, can help mitigate this issue. 
However, when the number of separated output nodes is limited, clustering methods are still required to assign separated streams to speakers for long recordings. 

In sequential cascade framework \cite{chen2020continuous, vinnikov24_interspeech, fang2021deep, von2019all, kinoshita2020tackling, boeddeker2018front, medennikov2020stc, niuISSD, niu2023qdm, boeddeker2024ts, taherian2024multi}, there is no definitive answer as to whether speech separation or speaker diarization should be performed first. Some approaches \cite{chen2020continuous, vinnikov24_interspeech, fang2021deep, von2019all, kinoshita2020tackling} choose to perform speech separation first, namely ``separation-then-diarization”. For example, the continuous speaker separation (CSS) pipeline \cite{chen2020continuous, vinnikov24_interspeech} first processes overlapping regions in the original audios through speech separation and then applies the separated results to diarization. Some studies \cite{fang2021deep, von2019all, kinoshita2020tackling} explore speaker diarization by performing VAD on the separated outputs, such as speech separation guided diarization (SSGD) \cite{fang2021deep} and recurrent selective attention network (RSAN) \cite{von2019all, kinoshita2020tackling}. 
Some methods \cite{boeddeker2018front, medennikov2020stc, niuISSD, niu2023qdm, boeddeker2024ts, taherian2024multi} begin with speaker diarization and then use the obtained time boundary information to guide speech separation, namely ``diarization-then-separation". This is also a promising direction, as speaker diarization is generally simpler than speech separation, allowing the former to provide a more accurate prior. A classic method is guided source separation (GSS) \cite{boeddeker2018front, medennikov2020stc}, which uses time boundaries to initialize time-varying mixture weights based on traditional spatial mixture models \cite{vu2010blind, ito2016complex}. However, GSS relies on multi-channel spatial information. 
The speaker activity driven speech extraction neural network (ADEnet) \cite{delcroix2021speaker} explores how to use speaker activity boundaries as auxiliary cues for speech extraction.
The CSS-AD \cite{von2024meeting} pipeline performs speech recognition on CSS outputs, followed by diarization on the ASR output, achieving promising results on the LibriCSS dataset \cite{chen2020continuous}.
Certain approaches \cite{niuISSD, niu2023qdm} enhance separation performance in real two-speaker scenarios by using speaker diarization to generate adaptation training data.
Recently, the target-speaker separation (TS-SEP) \cite{boeddeker2024ts} and speaker separation via neural diarization (SSND) \cite{taherian2024multi} have achieved strong performance in the fields of diarization and separation.
TS-SEP method combines speaker diarization and speech separation by extending the output of target-speaker voice activity detection (TS-VAD) \cite{Medennikov2020TargetSpeakerVA} from the time domain to the time-frequency domain. SSND uses estimated speaker boundaries to assign speakers to the outputs from a multi-speaker separation model. These two methods demonstrate the great potential of the sequential cascade framework. 

From various studies mentioned above, it is apparent that both joint training and sequential cascade approaches have advantages and limitations. Joint training allows simultaneous diarization and separation within a compact and unified framework. However, typical multi-task learning frameworks usually do not fully leverage diarization results to optimize separation results. Additionally, joint training frameworks are generally restricted to simulated data. Conversely, the sequential cascade approach—particularly ``diarization-then-separation"—allows for more effective integration of two tasks, where separation can effectively leverage diarization results. However, this approach typically involves a more cumbersome process and independently optimized components. 
Therefore, an effective approach is to combine ``joint training'' and ``sequential cascade", thereby taking advantage of both methods—what we refer to as ``deep cascade fusion". However, due to the differences in diarization and separation tasks, directly integrating them introduces a problem: diarization models can have many output nodes (e.g., eight output nodes in \cite{he2021target}). Directly feeding these outputs into the separation model and aligning the number of separation outputs with those of the diarization module may lead to non-convergence \cite{boeddeker2024ts}.  This issue arises because the separation model usually allocates an output node for each speaker. However, when training on typical conversational data, each node has limited target signals, making it difficult for the neural network to learn speaker-discriminative features. For example, in a natural eight-speaker meeting conversation, there are many regions where only one or two people are speaking, a separation model with eight outputs will have only one or two nodes with effective training targets in those regions. This significantly increases the training difficulty, which we term the sparse data convergence instability (SDCI) problem. This issue was also highlighted when the TS-SEP method was proposed \cite{boeddeker2024ts}, where a two-stage training approach was proposed as the solution. However, two-stage training makes the TS-SEP method unable to fully leverage the boundary information from the diarization module for separation. The SSND addresses this issue by constructing dual embedding sequences, which leads to a  complex framework. This approach requires a pre-trained diarization model to extract speaker embeddings, followed by constructing an embedding sequence based on boundary information during both the  training and testing phases. Additionally, a scenario-aware differentiated loss \cite{ao2023used, pan2022usev} has been proposed to handle the sparsely overlapped speech. However, this approach requires defining multiple different loss functions. 


To address the above issues, we propose a single-channel deep cascade fusion of diarization and separation (DCF-DS) framework for back-end multi-speaker speech recognition. The main contributions of our study are as follows:

\begin{enumerate}[label=\arabic*)]
	\item In DCF-DS, a joint training framework sequentially combining neural speaker diarization and separation is designed. Compared to TS-SEP, DCF-DS effectively leverages temporal boundaries from the diarization, reducing the difficulty of separation. Compared to SSND, DCF-DS avoids complex pre-processes and eliminates dependence on spatial information. 
	To complement DCF-DS training, we design a window-level decoding scheme to handle the SDCI problem. This decoding scheme allows the separation module to produce fewer outputs than the global number of speakers in an utterance.
	\item We introduce an optional multi-input multi-output speech enhancement module (MIMO-SE) within the DCF-DS framework, which directly utilizes the same structure as the separation module in DCF-DS, providing additional performance gains.
	\item We further optimize the speaker diarization results by re-clustering the outputs of DCF-DS, effectively enhancing back-end speech recognition performance. This process is typically not incorporated by advanced methods such as TS-SEP and SSND.
	\item By using the DCF-DS approach, we achieved the first place in the realistic single-channel track of the CHiME-8 NOTSOFAR-1 (Natural Office Talkers in Settings Of Far-field Audio Recordings) challenge \cite{vinnikov24_interspeech}. Meanwhile, our evaluation on the open-source LibriCSS dataset \cite{chen2020continuous} demonstrates state-of-the-art performance for single-channel speech recognition.
	
\end{enumerate}

The remainder of this paper is organized as follows. Section \ref{section2: prior works} provides an overview of related prior works. Section \ref{section3: DCF-DS} describes the proposed DCF-DS framework in detail. In Section \ref{section4: EX}, we present and analyze the experimental results. Finally, we conclude the paper in Section \ref{section5: Conclusion}.


\section{Prior Work}
\label{section2: prior works}
\subsection{Neural Speaker Diarization}
\label{subsection21: NSD}
\begin{figure}[t]
	\centering
	\includegraphics[width=0.7\linewidth]{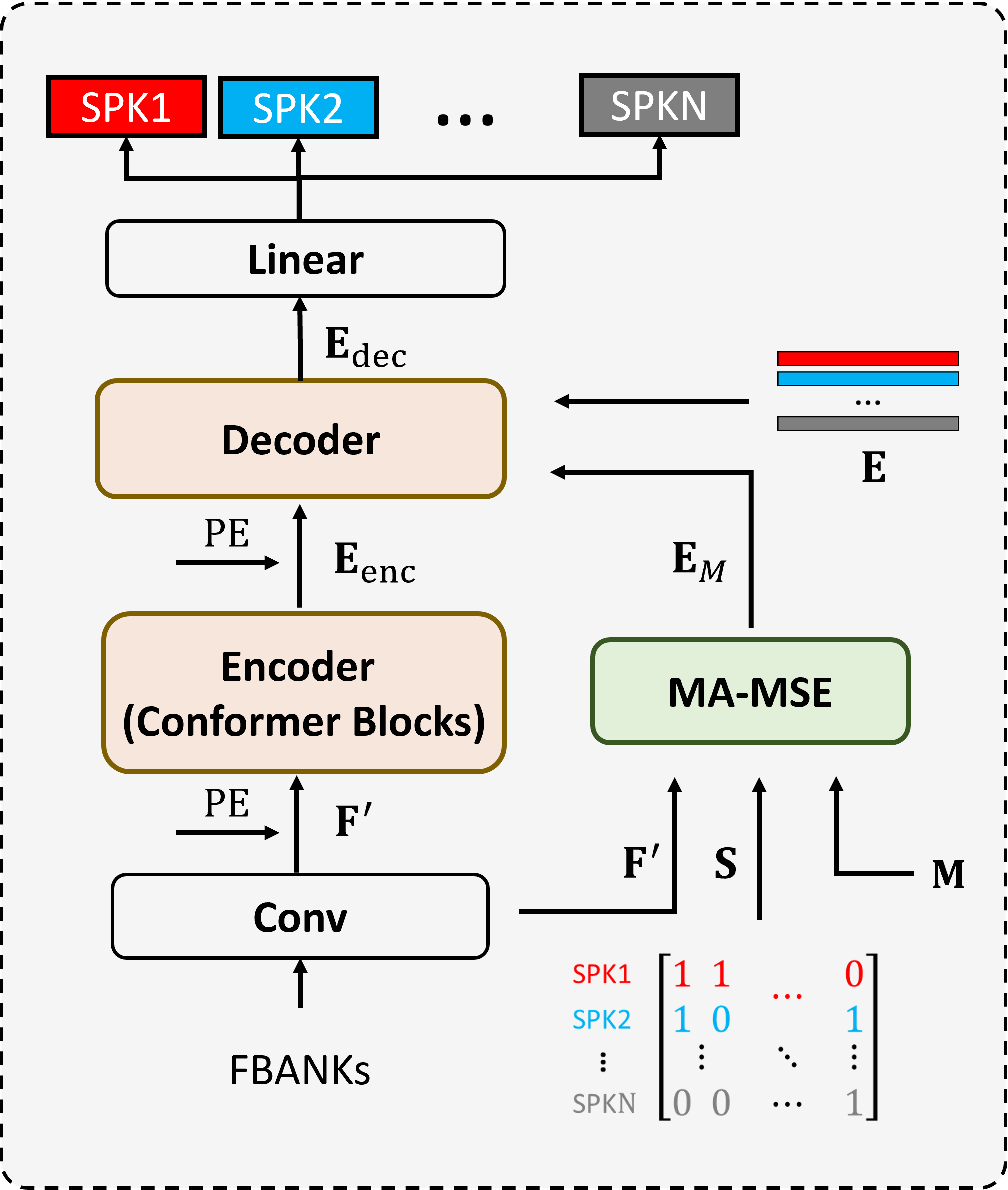}
	\caption{The overall architecture of NSD-MS2S.}
	\label{fig2}
	\vspace{-0.2cm}
\end{figure}

Traditional speaker diarization systems are mostly based on clustering methods \cite{AHC, diarizationLSTM, meanshif, dimitriadis2017developing, sell2018diarization, VBxdihard}. However, a limitation of these methods is their inability to effectively handle overlapping speech segments. 
To address this issue, end-to-end neural diarization techniques have been introduced, reformulating the speaker diarization task as a multi-target classification problem. For instance, end-to-end neural speaker diarization (EEND) \cite{fujita2020end, eenddea} simultaneously predicts all speakers’ activities on each frame. Target-speaker voice activity detection (TS-VAD) \cite{TSVAD, wang2021dku, he2021target} estimates the probability of speaker presence by using pre-enrolled speaker embeddings as auxiliary inputs.
In CHiME-7 DASR challenge \cite{cornell23_chime}, we proposed a neural speaker diarization system using memory-aware multi-speaker embedding with sequence-to-sequence architecture (NSD-MS2S) \cite{yang2024neural} based on Seq2Seq-TSVAD \cite{cheng2023target}, which demonstrates a significant improvement over the official baseline system. 
The overall architecture of NSD-MS2S is shown in Fig. \ref{fig2}. The inputs of NSD-MS2S are defined as follows:

\begin{equation}\label{eq1}
	\mathbf{X} = \{ \mathbf{x}_t \in \mathbb{R}^D \mid t = 1, \dots, T \}
\end{equation}
\begin{equation}\label{eq2}
	\mathbf{E} = \{ \mathbf{e}_n \in \mathbb{R}^L \mid n = 1, \dots, N \}
\end{equation}
\begin{equation}\label{eq3}
	\mathbf{S} = \{ s_{n,t} \in \{0, 1\} \mid n = 1, \dots, N ; t = 1, \dots, T \}
\end{equation}
where $\mathbf{X}$ represents the input acoustic features. In NSD-MS2S, we use log-Mel filter-bank features (FBANKs). $\mathbf{x}_t$ denotes the $D$-dimensional feature vector at frame $t$, where $T$ represents the total frame number. $ \mathbf{e}_n $ represents the speaker embedding of the $n$-th speaker, and $N$ denotes the number of speakers. $\mathbf{S}$ is the speaker mask matrix, which has elements $s_{n,t}$ that indicate whether speaker $n$ is present at frame $t$, taking a value of 1 if present and 0 if not. The speaker mask matrix $ \mathbf{S} $ is obtained through manual annotations during training and an initialized diarization system (e.g., spectral clustering) during testing.
The NSD-MS2S transforms the inputs into deep features through the following process:
\begin{equation}\label{eq4}
\mathbf{F} = \text{Conv}(\mathbf{X}) \in \mathbb{R}^{C \times T \times \frac{D}{2}}
\end{equation}
\begin{equation}\label{eq5}
\mathbf{F'} = \text{Downsample}(\mathbf{F}) \in \mathbb{R}^{T \times H}
\end{equation}
\begin{equation}\label{eq6}
\mathbf{E}_{M} = \text{MA-MSE}(\mathbf{F'}, \mathbf{S}, \mathbf{M}) \in \mathbb{R}^{N \times H_M}
\end{equation}
where $\text{Conv}(\cdot)$ denotes the convolutional layers, $C$ is the number of output channels. $ H $ is the feature dimension after downsampling.  $\text{MA-MSE}(\cdot)$ represents the memory-aware multi-speaker embedding (MA-MSE)  module proposed in \cite{he2023ansd}, which can provide dynamically refined speaker embeddings using the speaker embedding basis matrix $\mathbf{M}$. Next, NSD-MS2S passes these features through an encoder-decoder architecture to generate the final output:
\begin{equation}\label{eq7}
\mathbf{E}_{\text{enc}} = \text{Encoder}(\mathbf{F'} + \text{PE}) \in \mathbb{R}^{T \times H}
\end{equation}
\begin{equation}\label{eq8}
\mathbf{E}_{\text{dec}} = \text{Decoder}(\mathbf{E}, \mathbf{E}_{M}, \mathbf{E}_{D}, \mathbf{E}_{\text{enc}} + \text{PE}) \in \mathbb{R}^{N \times {H}}
\end{equation}
\begin{equation}\label{eq9}
\hat{\mathbf{Y}} = \sigma(\text{Linear}(\mathbf{E}_{\text{dec}})) \in \mathbb{R}^{N \times T}
\end{equation}
where $\mathbf{E}_{D}$ is decoder embedding initialized by a zero matrix. $\text{PE}$ represents the positional embedding. $\sigma$ means Sigmoid function. $\hat{\mathbf{Y}}$ is the final posterior probability matrix, with element $\hat{y}_{n,t}$ represents the probability of activity for speaker $n$ at frame $t$.
\subsection{Continuous Speech Separation}
\label{subsection22: CSS}
Continuous speech separation (CSS) was developed to address overlapping speech in natural conversations \cite{chen2020continuous, chen2021continuous}. In this paper, we adopt a Conformer-based speech separation model \cite{chen2021continuous}, consistent with the CSS module in CHiME-8 NOTSOFAR-1 baseline \cite{vinnikov24_interspeech}. For Conformer-based CSS, the inputs are defined as:
\begin{equation}\label{eq10}
	S(t, f) = |\text{STFT}(x(\tau))|
\end{equation}
where $ S(t, f) $ represents the magnitude of the Short-Time Fourier Transform (STFT) feature for the input time-domain signal $x(\tau)$. $t$ and $f$ are indices for the time frame and frequency, respectively. The separation process proceeds as follows:
\begin{equation}\label{eq11}
	\hat{M}_k(t, f) = \text{Conformer}(S(t, f)), \quad k = 1, \dots, K
\end{equation}
where $\hat{M}_k(t, f)$ denotes the T-F mask for speaker $k$. $K$ denotes the number of outputs. In traditional CSS \cite{chen2021continuous, vinnikov24_interspeech}, $K$ is typically set to 3, meaning the model can output up to three separated streams. This configuration constrains the length of the input signals (e.g., input segments are limited to 2.4 seconds in \cite{chen2021continuous} and 3 seconds in \cite{vinnikov24_interspeech}). To handle this limitation, chunk-wise decoding is commonly applied, with overlapping regions between adjacent chunks used to align the $K$ separated streams. In processing multi-speaker utterances, the CSS-separated results also require corresponding speaker attribution. One approach is to cluster the separated audios to determine the speaker identity for each segment \cite{xiao2021microsoft}. The NOTSOFAR-1 baseline \cite{vinnikov24_interspeech} utilizes the Whisper “large-v3” \cite{radford2023robust} model to transcribe the CSS-separated results, providing both the text and word-level timestamps. These timestamped segments are then assigned to specific speakers based on spectral clustering (SC), namely the ``CSS + SC'' framework.

\begin{figure*}[t]
	\centering
	\includegraphics[width=1\linewidth]{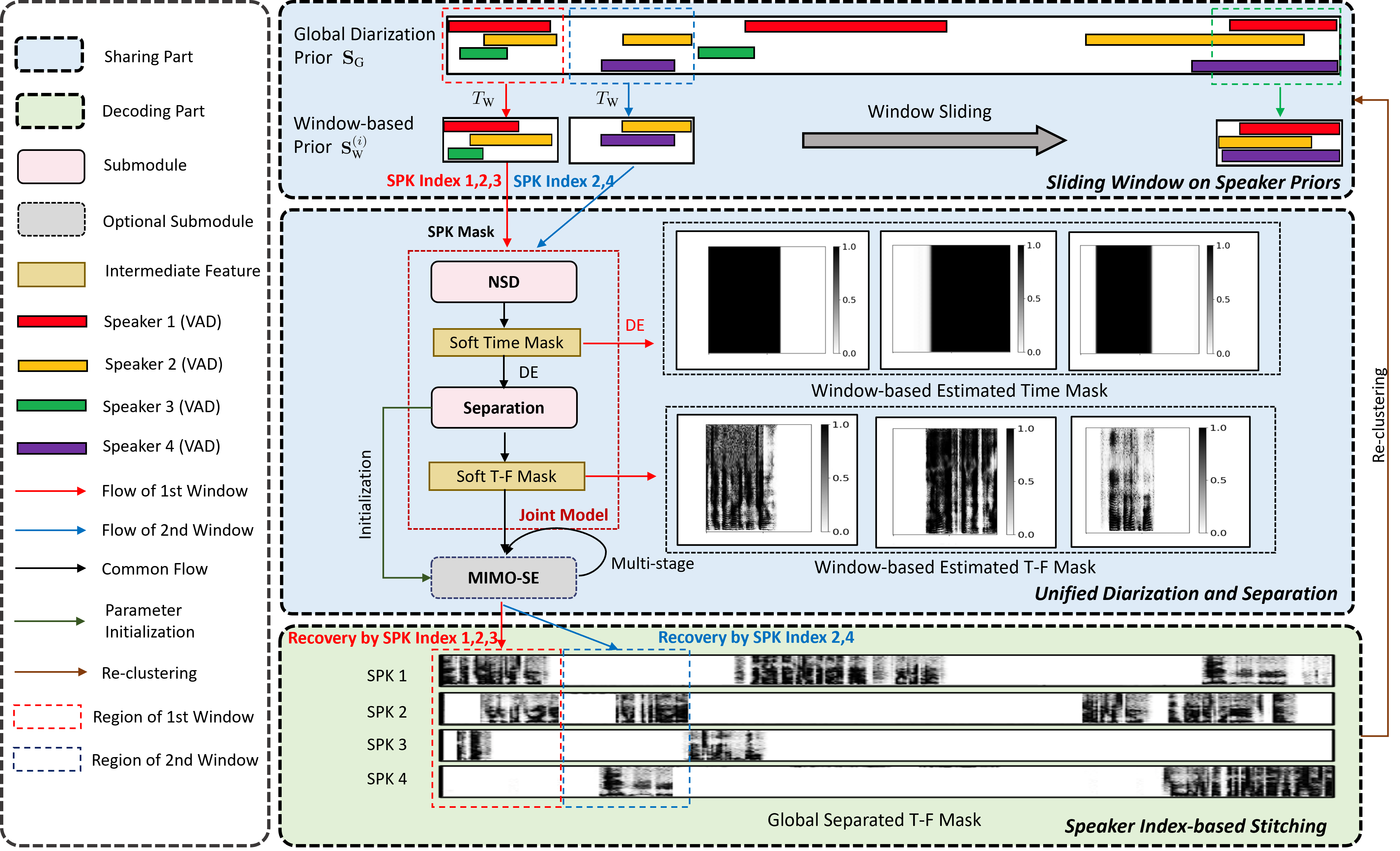}
	\caption{Overall framework of the proposed DCF-DS. ``DE'' means dimension extension. The diagram illustrates the processing of an example utterance involving 4 speakers. As a demonstration, the DCF-DS framework employs a 3-second sliding window ($T_{\text{W}}=3$) and produces 3 separated streams, consistent with the NOTSOFAR-1 CSS baseline \cite{vinnikov24_interspeech} (other configurations are also explored in our experiments). Acoustic feature inputs, such as FBANK features, are omitted due to space constraints.}
	\label{fig3}
\end{figure*}
\section{Deep Cascade Fusion of Diarization and Separation Framework}
\label{section3: DCF-DS}
Fig. \ref{fig3} illustrates the overall DCF-DS framework, which is composed of three main components: sliding window on speaker priors (SWSP), unified diarization and separation (UDS), and speaker index-based stitching (SIS). The first two components are utilized during both training and decoding, while the SIS component is employed exclusively during decoding to generate globally separated streams that are consistent with the speaker identities in the global context.
Similar to the NSD-MS2S system, DCF-DS also requires an initial speaker diarization result, which may include more speakers than the number of separated outputs in the DCF-DS (for example, as shown in the Fig. \ref{fig3}, there are 4 global speakers, but only 3 separated output streams). To handle this, the SWSP process converts the original global speaker distribution into a windowed distribution, ensuring that the number of speakers within each window is generally less than or equal to the number of output nodes. This allows the window-based speaker priors to be fed into the UDS component, which performs window-based diarization and separation. Finally, the SIS component uses global indices to stitch together the window-based separation results in the decoding phase, yielding the complete global speech separation outputs, as depicted at the bottom of Fig. \ref{fig3}. The details of each module will be described in the following sections.

\subsection{Sliding Window on Speaker Priors}
\label{section31: SWSP}

The global diarization prior for a multi-speaker time-domain audio $s(\tau)$ can be represented by a speaker mask matrix:

\begin{equation}\label{eq12}
\mathbf{S}_{\text{G}} = \{ s_{n_{\text{G}}, t_{\text{G}}} \in \{0, 1\} \mid n_{\text{G}} = 1, \ldots, N_{\text{G}}; \, t_{\text{G}} = 1, \ldots, T_{\text{G}} \}
\end{equation}
where $s_{n_{\text{G}}, t_{\text{G}}}$ indicates whether the $n_{\text{G}}$-th speaker is active at frame $t_{\text{G}}$, with a value of 1 representing presence and 0 representing absence. $ n_{\text{G}} $ and $ t_{\text{G}} $ are global indices. $N_{\text{G}}$ denotes the total number of speakers in $s(\tau)$. $ T_{\text{G}}$ represents the total frame number of the entire utterance. 
To manage the large number of speakers and the long duration of $s(\tau)$, we utilize the sliding window on speaker priors (SWSP) module. This module segments the global speaker priors into smaller, window-based segments using a fixed-length sliding window. The window-based prior is defined as:

\begin{equation}\label{eq13}
\mathbf{S}_{\text{W}}^{(i)} = \{ s_{n_{\text{W}}, t_{\text{G}}} \in \{0, 1\} \mid 1 \leq n_{\text{W}} \leq N_{\text{W}}; \, t_{\text{G}} = t_i, \ldots, t_i + L - 1 \}
\end{equation}
where $\mathbf{S}_{\text{W}}^{(i)}$ denotes the speaker priors within the $i$-th window. $n_{\text{W}} \in \mathbb{Z}^+$ indicates the speaker index within the window. $t_i$ is the starting frame index of the $i$-th window. $L$ is the fixed window frame length. $N_{\text{W}}$ represents the maximum speaker number within a window. 

During the SWSP process, we compact the speaker index within each window. For example, in the second window at the top of Fig. \ref{fig3}, we convert the global indices $ n_{\text{G}} = 2, 4 $ into the windowed indices $ n_{\text{W}} = 1, 2 $, effectively controlling the number of speakers $ N_{\text{W}} $ within the window. We also save this correspondence using the mapping function $ f_{\text{map}}^{(i)} : n_{\text{W}} \rightarrow n_{\text{G}} $ for the $ i $-th window, which will be used in the subsequent decoding process. 
In rare cases during decoding, if the number of speakers within a window exceeds $ N_{\text{W}} $, we select the $ N_{\text{W}} $ speakers with the longest duration in the priors. During training, such windows are discarded and not used for training. Conversely, if $ N_{\text{W}} $ is greater than the number of speakers in the current window, we apply the zero-padding. 
Note that we use manually annotated global diarization priors during training, while in decoding, we employ an auxiliary speaker diarization system to provide the global segmentation priors as in NSD-MS2S.
The SWSP approach ensures that the number of separation output nodes $N_{\text{W}}$ remains not too large, which allows each separation node to be sufficiently activated during training. During decoding, the SWSP approach enables a limited number of separation output nodes to handle an arbitrary number of global speakers while preserving crucial long context information in $\mathbf{S}_{\text{G}}$. 
 
\subsection{Unified Diarization and Separation}
\label{section32: UDS}
\begin{figure}[t]
	\centering
	\includegraphics[width=1.0\linewidth]{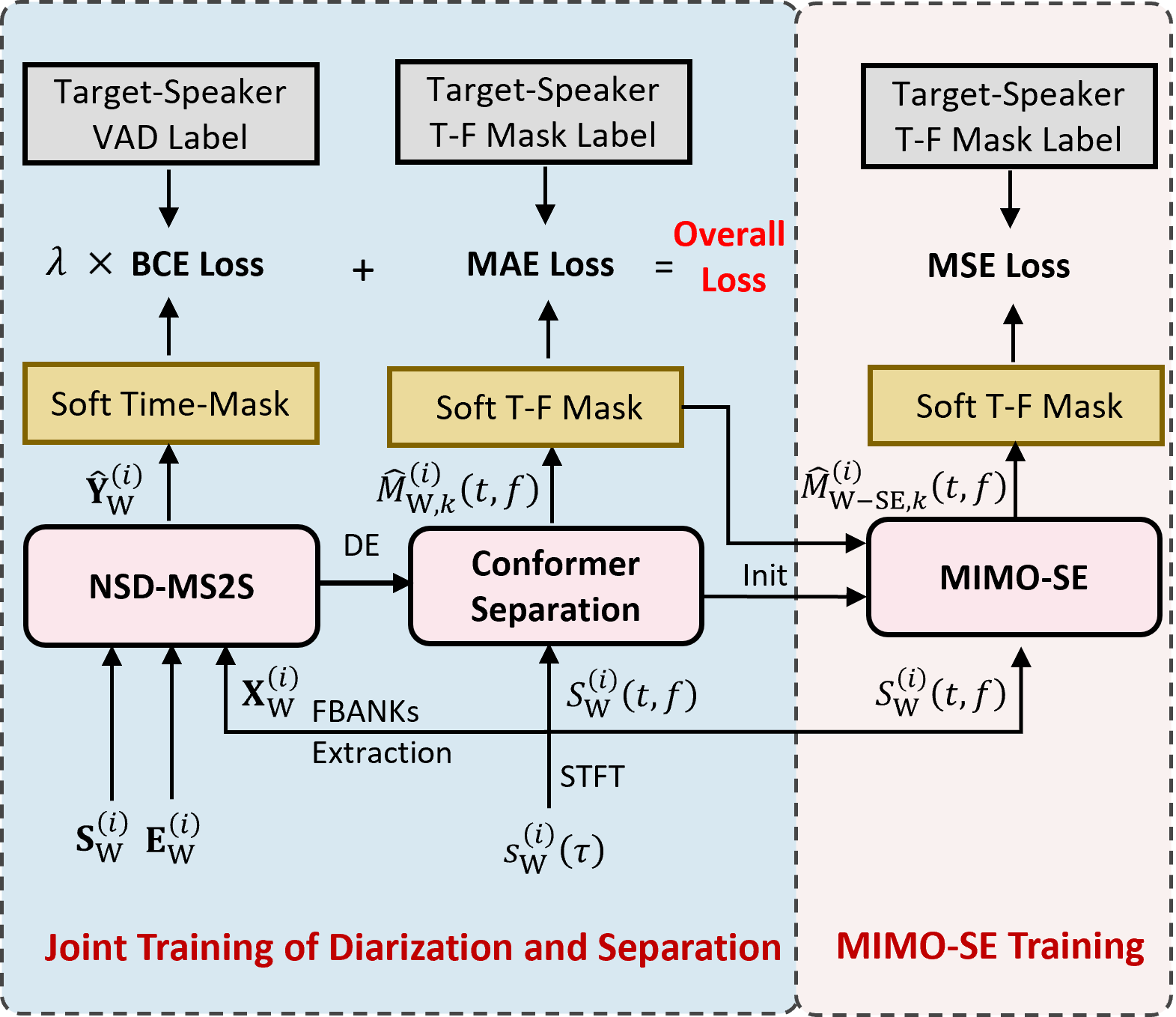}
	\vspace{-0.5cm}
	\caption{The training process of DCF-DS framework. ``DE'' means dimension extension.}
	\label{fig4}
	\vspace{-0.2cm}
\end{figure}
Once the windowed prior $ \mathbf{S}_{\text{W}}^{(i)} $ is obtained, it can be used to perform joint diarization and separation, as illustrated in Fig. \ref{fig3}. This module contains the core neural network component of the DCF-DS framework, with its training process illustrated in Fig. \ref{fig4}. The training process involves two main workflows: the primary joint training process and an optional MIMO-SE training process. 
In the first part, we input the window-based speaker embeddings $ \mathbf{E}_{\text{W}}^{(i)} $ (corresponding to the speakers within the $ i $-th window) and the window-based FBANKs $ \mathbf{X}_{\text{W}}^{(i)} $ into the sequential joint model. First, these features are fed into the NSD-MS2S architecture as shown in Fig. \ref{fig4}. The process can be described as:

\begin{equation}\label{eq14}
	\hat{\mathbf{Y}}_{\text{W}}^{(i)} = \text{NSD-MS2S}(\mathbf{S}_{\text{W}}^{(i)}, \mathbf{E}_{\text{W}}^{(i)}, \mathbf{X}_{\text{W}}^{(i)}) \in \mathbb{R}^{N_{\text{W}} \times L}
\end{equation}
where $ \hat{\mathbf{Y}}_{\text{W}}^{(i)}$ represents the probability of speaker presence for $ N_{\text{W}} $ speakers across $ L $ frames within the $ i $-th window. After obtaining $ \hat{\mathbf{Y}}_{\text{W}}^{(i)} $, we utilize the time boundary information it provides to enhance the performance of the speech separation module. Assume that the time-domain signal within the $i$-th window is denoted as $ s_{\text{W}}^{(i)}(\tau) $, we can obtain the STFT magnitude feature as:

\begin{equation}\label{eq15}
	S_{\text{W}}^{(i)}(t, f) = \left| \text{STFT}\left( s_{\text{W}}^{(i)}(\tau) \right) \right| \in \mathbb{R}^{L \times F}
\end{equation}

We then perform a dimensional extension (DE) by replicating $ \hat{\mathbf{Y}}_{\text{W}}^{(i)} $ along an additional axis, which is treated as the frequency dimension. This operation can be expressed as:

\begin{equation}\label{eq16}
	\hat{\mathbf{Y}}_{\text{W-DE}}^{(i)} = \text{Broadcast}(\hat{\mathbf{Y}}_{\text{W}}^{(i)}) \in \mathbb{R}^{N_{\text{W}} \times L \times F}
\end{equation}

The added dimension ensures that each speaker's time mask in $ \hat{\mathbf{Y}}_{\text{W-DE}}^{(i)} $ matches the shape of the STFT magnitude feature as shown in Fig. \ref{fig3}, enabling the combined features to be processed by the separation model:

\begin{equation}\label{eq17}
	\mathbf{I}_{\text{concat}}^{(i)} = \text{Concat}(\hat{\mathbf{Y}}_{\text{W-DE}}^{(i)}, S_{\text{W}}^{(i)}(t, f)) \in \mathbb{R}^{(N_{\text{W}} + 1) \times L \times F}
\end{equation}
\vspace{-0.05cm}
\begin{equation}\label{eq18}
	\hat{M}_{\text{W}, n_{\text{W}}}^{(i)}(t, f) = \text{Conformer}(\mathbf{I}_{\text{concat}}^{(i)}), \quad n_{\text{W}} = 1, \dots, N_{\text{W}}
\end{equation}

The concatenated features combine spectral information and temporal boundary information into a unified input. This integration effectively supplies the model with speaker identity, order, and temporal boundary priors, mitigating the permutation problem \cite{kolbaek2017multitalker}. Additionally, comparing Eqs. \eqref{eq11} and \eqref{eq18}, we can see that this unified input changes the nature of the separation task: instead of mapping directly from the mixed speech signals, the model now maps from the time-frame distribution of speakers to their time-frequency distribution. This adjustment simplifies the speech separation task compared to the traditional approach. 
After the separation model, we introduce an optional MIMO-SE module, which directly utilizes the Conformer-based speech separation outputs in Eq. \eqref{eq18} to generate the cleaner T-F mask for each speaker. The MIMO-SE takes the combination of the separation results and the STFT magnitude features as input:

\begin{equation}\label{eq19}
	\mathbf{I}_{\text{SE-concat}}^{(i)} = \text{Concat}(\hat{M}_{\text{W}, 1:N_{\text{W}}}^{(i)}(t, f), S_{\text{W}}^{(i)}(t, f))
\end{equation}
where $\mathbf{I}_{\text{SE-concat}}^{(i)} \in \mathbb{R}^{(N_{\text{W}} + 1) \times L \times F}$. The MIMO-SE module uses the Conformer architecture to process these concatenated features, producing the enhanced T-F mask for each speaker:

\begin{equation}\label{eq20}
	\hat{M}_{\text{W-SE}, n_{\text{W}}}^{(i)}(t, f) = \text{Conformer}(\mathbf{I}_{\text{SE-concat}}^{(i)})
\end{equation}
where $n_{\text{W}} = 1, \dots, N_{\text{W}}$.
The input and output dimensions in Eq. \eqref{eq20} are consistent with those of the Conformer-based speech separation model in Eq. \eqref{eq18}, allowing for the reuse of the same Conformer architecture. By comparing Eqs. \eqref{eq11}, \eqref{eq18}, and \eqref{eq20}, we can see that MIMO-SE serves as an extension of the separation module, simplifying the task further: rather than mapping from a T-mask to a T-F mask as in Eq. \eqref{eq18}, it now maps from a T-F mask to a cleaner T-F mask.
The enhancement module aims to refine the quality of the separated speech by utilizing the initial separation results, ultimately producing more robust and cleaner separated streams.

During training, we sequentially cascade the diarization and separation modules for joint training, as shown in Fig. \ref{fig4}. In the joint training process, we apply binary cross-entropy (BCE) loss to the diarization module:

\begin{equation}\label{eq21}
	\mathcal{L}_{\text{BCE}} = \frac{1}{L N_{\text{W}}} \sum_{n=1}^{N_{\text{W}}} \sum_{t=1}^{L} \text{BCE}(\hat{y}_{n, t}^{(i)}, y_{n, t}^{(i)})
\end{equation}
where $ \hat{y}_{n, t}^{(i)} $ and $ y_{n, t}^{(i)} $ are elements of the matrices $ \hat{\mathbf{Y}}_{\text{W}}^{(i)} $ and $ \mathbf{Y}_{\text{W}}^{(i)} $, respectively, representing the predicted present probabilities and ground truth labels for speaker $ n $ at frame $ t $ within the $ i $-th window. 
For the T-F masks $ \hat{M}_{\text{W}, n_{\text{W}}}^{(i)}(t, f) $ estimated by the separation module, we use mean absolute error (MAE) loss:

\begin{equation}\label{eq22}
	\mathcal{L}_{\text{MAE}} = \frac{1}{L F N_{\text{W}}} \sum_{n_{\text{W}}=1}^{N_{\text{W}}} \sum_{t=1}^{L} \sum_{f=1}^{F} \left|  \hat{M}_{\text{W}, n_{\text{W}}}^{(i)}(t, f) - M_{\text{W}, n_{\text{W}}}^{(i)}(t, f) \right|
\end{equation}
where $ M_{\text{W}, n_{\text{W}}}^{(i)}(t, f) $ represents the ground truth T-F mask for speech separation defined as:

\begin{equation}\label{eq23}
	M_{\text{W}, n_{\text{W}}}^{(i)}(t, f) = \frac{ S_{\text{W}, n_{\text{W}}}^{(i)}(t, f) }{ S_{\text{W}}^{(i)}(t, f) }
\end{equation}
where $ S_{\text{W}, n_{\text{W}}}^{(i)}(t, f) $ is the magnitude of the clean speech for speaker $ n_{\text{W}} $ under the $i$-th window.
The overall loss for joint training is a weighted combination of the two losses:

\begin{equation}\label{eq24}
	\mathcal{L}_{\text{overall}} = \lambda \times \mathcal{L}_{\text{BCE}} + \mathcal{L}_{\text{MAE}}
\end{equation}
where $ \lambda $ controls the weight of the BCE loss, allowing us to balance the contributions of the diarization and separation tasks during training. In the overall training process, we optimize both the diarization and the separation components. This is because training solely with the separation loss would reduce the entire joint model to merely a large speech separation model, which defeats the purpose of joint training. Introducing the diarization loss $ \mathcal{L}_{\text{BCE}} $ ensures that the diarization module continues to provide time boundary information. Additionally, the predicted probabilities of the diarization module may contain unavoidable errors. The joint training process prevents the speech separation module from over-fitting to these predictions, which constantly updates the diarization model's parameters using the correct labels and maintains the robustness of the entire system.

After the joint training, we can use the same loss function as in Eq. \eqref{eq22} for the optional MIMO-SE module:

\begin{equation}\label{eq25}
	\mathcal{L}_{\text{MAE}}^{'} = \frac{1}{L F N_{\text{W}}} \sum_{k=1}^{N_{\text{W}}} \sum_{t=1}^{L} \sum_{f=1}^{F} \left| \hat{M}_{\text{W-SE}, k}^{(i)}(t, f) - M_{\text{W}, k}^{(i)}(t, f) \right|
\end{equation}

Since the MIMO-SE model structure is identical to the separation component in the joint training process, we can initialize the MIMO-SE module with the parameters from the trained separation module, which helps accelerate convergence, as shown in Fig. \ref{fig4}. 
\subsection{Speaker Index-based Stitching}
\label{section33: SIS}
After the UDS process, we obtain the separation results for each window. It is necessary to reconstruct the global separation results from these windowed outputs during the  decoding phase. Using the mapping function $ f_{\text{map}}^{(i)} : n_{\text{W}} \rightarrow n_{\text{G}} $ saved during the SWSP process, we restore the windowed separation result to form the global separation output:
\begin{equation}\label{eq26}
	\hat{M}_{\text{G}, n_{\text{G}}}^{(i)}(t_G, f) = \sum_{n_{\text{W}}=1}^{N_{\text{W}}}\mathbb{I}(n_{\text{G}} =  f_{\text{map}}^{(i)}(n_{\text{W}})) \cdot \hat{M}_{\text{W}, n_{\text{W}}}^{(i)}(t, f)
\end{equation}
where $ \hat{M}_{\text{G}, n_{\text{G}}}^{(i)}(t_G, f) $ represents the global T-F mask for speaker $ n_{\text{G}} $ in prior $\mathbf{S}_{\text{G}}$ under the $i$-th window. The global frame index $t_G$, corresponding to Eq. \eqref{eq13}, ranges from $t_i$ to $t_i + L - 1$. The indicator function $ \mathbb{I} $ checks if the global speaker index $ n_{\text{G}} $ corresponds to the windowed speaker index $ n_{\text{W}} $ via the mapping function. 
The separation result $\hat{M}_{\text{W}, n_{\text{W}}}^{(i)}(t, f) $ used here can also be substituted with the output from the MIMO-SE module, providing a refined T-F mask. This approach enables seamless reconstruction of the separated streams for all speakers across the entire recording.
After obtaining the global prediction across all windows, we can derive the STFT of the reconstructed separated signal:
\begin{equation}\label{eq27}
	\hat{S}_{\text{G}, n_{\text{G}}}(t_G, f) = \hat{M}_{\text{G}, n_{\text{G}}}(t_G, f) \cdot Z(t_G, f)
\end{equation}
where $\hat{M}_{\text{G}, n_{\text{G}}}(t_G, f)$ and $Z(t_G, f)$ represent the estimated global T-F mask and the STFT of the mixed signal (including phase information) for the entire utterance, respectively. The frame index $t_G$ ranges from 1 to $T_{\text{G}}$, where $T_{\text{G}}$ is the total frame number in the utterance as defined in Eq. \eqref{eq12}.
The global separated time-domain signal $\hat{s}_{n_{\text{G}}}(\tau)$ is obtained by applying the inverse STFT to $\hat{S}_{\text{G}, n_{\text{G}}}(t_G, f)$. After obtaining the separated signals for each speaker, re-clustering is performed to obtain an enhanced global diarization prior:
\begin{equation}\label{eq28}
	\mathbf{S}_{\text{G}} = \text{Re-cluster}\left(\{\hat{s}_{n_{\text{G}}}(\tau) \mid n_{\text{G}} = 1, \ldots, N_{\text{G}}\}\right)
\end{equation}

This re-clustering process is illustrated on the right side of Fig. \ref{fig3}. One of the main advantages of re-clustering is that it reduces speaker confusion errors. This is mainly due to the reduction of overlap regions in the separated speech, thus improving the reliability of the speaker clustering. By leveraging these improved diarization priors, the overall performance can be further enhanced. 

\section{Experiments and Result Analyses}
\label{section4: EX}
\subsection{Datasets}
\label{subsection41: DD}
Our primary focus is on real-world multi-speaker single-channel scenarios. Thus, we primarily evaluate the proposed method on the single-channel track of the NOTSOFAR-1 eval small dataset \cite{vinnikov24_interspeech}, consistent with the evaluation set used in the challenge, allowing direct comparison with other submitted systems. The NOTSOFAR-1 evaluation set comprises 80 real meeting recordings sampled at 16 kHz, each captured by two different single-channel devices. It consists of English conversations lasting approximately 6 minutes per session, involving 4 to 8 participants with balanced gender representation. 
The evaluation set covers a wide range of real-world complexities, such as  diverse speaker distances, varying volumes, multi-level background noises, and dynamic changes in the acoustic transfer function. It also includes conversational challenges like overlapping speech, rapid speaker transitions, interruptions, and so on. Collected across 10 rooms with distinct acoustic profiles, this dataset provides a well-established benchmark for multi-speaker speech recognition in realistic environment.
We use the development set in NOTSOFAR-1 dataset as our development set for hyper-parameter tuning. 
For the training set, we first utilize the official 1000-hour NOTSOFAR-1 simulated dataset, which employs the Image Method \cite{allen1979image} to simulate multi-channel data using clean speech from the Librivox corpus \cite{kearns2014librivox}. Since our focus is on training a single-channel system, we use the first channel for the training process. Additionally, we generate approximately 700 hours of simulated data using open-source scripts\footnote{https://github.com/jsalt2020-asrdiar/jsalt2020\_simulate} based on the near-field close-talking recordings corpus in released training and development datasets. These two simulated datasets are used for joint training in DCF-DS, as well as for training the MIMO-SE module. Furthermore, the NOTSOFAR-1 challenge organizers have released a set of 53-hour real-world meeting training dataset. Although these real recordings lack clean source signals and cannot be directly used for training the separation model, we leverage them, in combination with the simulated training data, to train an independent NSD-MS2S model, which can provide more accurate diarization priors for the DCF-DS framework. 

Additionally, to facilitate comparisons with more methods from the literature, we also evaluate our approach on the widely used LibriCSS corpus \cite{chen2020continuous}. The libriCSS dataset comprises 10 one-hour recordings. These recordings contain many utterances from LibriSpeech \cite{librispeech}, played back through loudspeakers in the meeting room, with overlap ratios ranging from 0$\%$ to 40$\%$. Each session involves 8 speakers. Consistent with previous studies \cite{boeddeker2024ts, chen2021continuous}, we adopt the single-channel configuration in the publicly available script\footnote{https://github.com/chenzhuo1011/libri\_css}. It is worth noting that we test our trained model directly on the LibriCSS dataset without applying any specific domain adaptation.

\subsection{Implementation Details}
\label{subsection42: ID}
We adopt the NOTSOFAR-1 challenge baseline system \cite{vinnikov24_interspeech} as the baseline system for our evaluation on the NOTSOFAR-1 eval set, following the ``CSS + SC'' pipeline as described in Section \ref{subsection22: CSS}.
In the DCF-DS framework, we use a consistent window length of 64 ms and a hop size of 16 ms for the STFT transformation. The NSD-MS2S module processes 40-dimensional FBANK features, with the encoder composed of 6 Conformer blocks with the same structure as in \cite{gulati20_interspeech}. The decoder module consists of 6 blocks with uniform settings: two attention layers with 512 dimensions and 8 attention heads, along with 1024-dimensional feed-forward layers. For the separation component, we adopt the Conformer architecture \cite{chen2021continuous} from the NOTSOFAR-1 baseline \cite{vinnikov24_interspeech} with one key modification: instead of using only the amplitude spectrum of the mixed speech as input, we include both the amplitude spectrum and soft time masks. Our Conformer configuration features an attention dimension of 512, 8 attention heads, and 18 blocks.
During training, we employ the Adam optimizer \cite{adam} with a learning rate of $ 1 \times 10^{-4} $ and conduct the training for 20 epochs (we also explore the performance with different numbers of epochs in preliminary experiments). The weight $ \lambda $ of the loss function in Eq. \eqref{eq24} is set to 1. We explore two different window length configurations: $ T_{\text{W}} = 3 $ seconds with the maximum speaker number $ N_{\text{W}} = 3 $ per window, aligning with the NOTSOFAR-1 baseline, and $ T_{\text{W}} = 12.8 $ seconds with the maximum speaker number $ N_{\text{W}} = 4 $, which enables the model to capture more context. To accelerate convergence, we initialize the diarization and separation modules in DCF-DS with pre-trained weights from NSD-MS2S and CSS trained on the simulated datasets. During the decoding phase, we use spectral clustering \cite{park2019auto} to generate the global diarization prior $ \mathbf{S}_{\text{G}} $. 
For the LibriCSS dataset, we utilize the speaker embedding extractor from the ESPnet LibriCSS recipe \cite{watanabe18_interspeech}. For the NOTSOFAR-1 dataset, we use the overlap detection module to assist the SC in handling overlapping regions as in \cite{niu24_chime}. The speaker embeddings are extracted using a ResNet-221 model \cite{niu24_chime} trained on the VoxCeleb \cite{nagrani2020voxceleb} and LibriSpeech datasets.  Additionally, as mentioned in Section \ref{subsection41: DD}, we independently train an NSD-MS2S system on both simulated and real training datasets to provide more accurate priors, employing the default configuration from the open-source code\footnote{https://github.com/liyunlongaaa/NSD-MS2S}.
For the NOTSOFAR-1 eval dataset, we utilize the Whisper ``large-v3" model \cite{radford2023robust} as the ASR back-end, consistent with the NOTSOFAR-1 challenge baseline \cite{vinnikov24_interspeech}. For the LibriCSS dataset, we use two pre-trained open-source models from the ESPnet framework \cite{watanabe18_interspeech}. The first model, which we refer to as E2E, is an end-to-end Transformer based on the LibriSpeech recipe\footnote{https://github.com/espnet/espnet/blob/master/egs/librispeech/asr1}, comprising a decoder with 6 self-attention blocks and an encoder with 12 blocks, trained on the 960-hour LibriSpeech dataset. The second model, which we refer to as E2E-SSL, leverages WavLM \cite{chen2022wavlm} self-supervised features (SSL) and uses a Conformer architecture\footnote{https://huggingface.co/espnet/simpleoier\_librispeech\_asr\_train\_asr\_\\conformer7\_wavlm\_large\_raw\_en\_bpe5000\_sp}, achieving a WER of 1.9$\%$ on the LibriSpeech clean test set.
\subsection{Performance Measures}
\label{subsection42: PM}
In our experiments, we primarily evaluate the impacts of different front-end methods on speech quality using the back-end ASR performance. For the LibriCSS dataset, we evaluate ASR performance using the concatenated minimum-permutation word error rate (cpWER) \cite{watanabe2020chime6}. cpWER is a widely used metric for multi-speaker scenarios which concatenates transcriptions for each speaker and finds the permutation that minimizes the WER.
For the NOTSOFAR-1 eval dataset, we adopt the time-constrained minimum-permutation word error rate (tcpWER) \cite{MeetEval23} as our evaluation metric, consistent with the NOTSOFAR-1 challenge. The tcpWER is an extension of cpWER that accounts for word errors, temporal boundary mismatches, and speaker identification errors. Even when words are correctly recognized, significant discrepancies in time boundaries result in penalties. Compared to cpWER, tcpWER incorporates temporal constraints and offers faster computation. In line with the official baseline in NOTSOFAR-1 challenge \cite{vinnikov24_interspeech}, we apply tcpWER with a 5-second collar.
Additionally, to further analyze the proposed method, we measure the diarization error rate (DER) \cite{DER} for certain systems, which includes miss (MI), false alarm (FA), and confusion (CF) errors. We do not apply the forgiveness collars to the DER calculation.
\subsection{Performance Comparison on NOTSOFAR-1}
\label{subsection43: NS}
\begin{table}[t]
	\renewcommand\arraystretch{1.25}
	\newcolumntype{L}[1]{>{\raggedright\arraybackslash}p{#1}}
	\newcolumntype{C}[1]{>{\centering\arraybackslash}p{#1}}
	\newcolumntype{R}[1]{>{\raggedleft\arraybackslash}p{#1}}
	\centering
	\caption{tcpWER ($\%$) of different front-end systems on the single-channel NOTSOFAR-1 eval-small dataset. For the back-end speech recognition system, we uniformly use the Whisper ``large-v3'' model \cite{radford2023robust}. $\mathbf{S}_{\text{G}}$ means the global diarization prior defined in Eq. \eqref{eq12}. ``RTM'' denotes replacing the  time mask. The sliding window length of DCF-DS is set to 3 seconds.}
	\label{tab:1}\medskip
	\resizebox{8.5 cm}{!}{\begin{tabular}{c|c|c|c}
			\toprule[1 pt]
			\textbf{Front-end Systems}&\textbf{Epoch}&$\mathbf{S}_{\text{G}}$&\textbf{tcpWER} ($\%$)\\
			\midrule
			Oracle Segment&-&Oracle Segment&39.73 \\
			NOTSOFAR-1 Baseline \cite{vinnikov24_interspeech} &-&-&41.4 \\
			\midrule
			\multirow{6}*{DCF-DS} &20&\multirow{3}*{SC}&38.28 \\
			&30&&38.05 \\
			&40&&\textbf{38.02} \\
			\cmidrule(r){2-4} 
			&20&\multirow{3}*{NSD-MS2S}&37.91 \\
			&30&&\textbf{37.66} \\
			&40&&37.80 \\
			\midrule
			\multirow{3}*{DCF-DS (with RTM)} &20&\multirow{3}*{NSD-MS2S}&37.31 \\
			&30&&37.26 \\
			&40&&\textbf{37.14} \\
			\bottomrule[1 pt]	
	\end{tabular}}
	\vspace{-0.3 cm}
\end{table}
Table \ref{tab:1} shows the performance of different front-end systems with the same ASR back-end. For DCF-DS, we set the window length $T_{\text{W}} = 3$ and the maximum speakers per window $N_{\text{W}} = 3$, consistent with the NOTSOFAR-1 baseline configuration. The training set for DCF-DS in this table is also restricted to the NOTSOFAR-1 simulated dataset, consistent with the baseline setup. From the table, we can observe that utilizing oracle diarization priors to segment the original mixed speech yields a tcpWER of 39.73$\%$. Even with oracle speaker boundaries, the segmented audios still contain substantial overlap regions, making it difficult to achieve high recognition accuracy and illustrating the challenges of the dataset. 
The NOTSOFAR-1 baseline system achieves a tcpWER of 41.4$\%$, slightly worse than the result obtained with oracle diarization priors. In the DCF-DS framework, using spectral clustering as the global prior results in a tcpWER of 38.02$\%$, representing an absolute improvement of 3.38$\%$ over the NOTSOFAR-1 baseline. We also compared performance across three different epochs (20, 30, and 40) and observed minimal differences among them. Furthermore, we incorporate an independently trained NSD-MS2S system, enhanced with real data, to provide the prior $ \mathbf{S}_{\text{G}} $. This improved prior leads to consistent performance gains across epochs 20, 30, and 40, reducing the best tcpWER from 38.02$\%$ to 37.66$\%$. These results indicate that a more accurate diarization prior enhances the DCF-DS results, which is consistent with our expectations.
Finally, we replace the outputs of the diarization module in DCF-DS, namely $ \hat{\mathbf{Y}}_{\text{W}}^{(i)} $ from Eq. \eqref{eq14}, with the probabilities predicted by the independently trained NSD-MS2S system. This means substituting the ``soft time mask" in Figures \ref{fig3} and \ref{fig4}, which we refer to as ``replace time mask (RTM)". An interesting observation emerges: even though this independently trained NSD-MS2S model was not included in the joint training process, incorporating its predicted probabilities directly into the separation module of DCF-DS resulted in performance improvements. This can be attributed to the addition of real data in the training set, indicating that the DCF-DS framework can effectively leverage real-world data to enhance final results. Since we observe minimal performance differences across epochs 20, 30, and 40, we consistently use the results from epoch 20 in subsequent experiments for training efficiency. Moreover, among the different DCF-DS configurations, using RTM yields the best performance. Therefore, unless otherwise specified, all subsequent experiments use DCF-DS with RTM as the default configuration.
\begin{figure*}[t]
	\centering
	\includegraphics[width=0.95\linewidth]{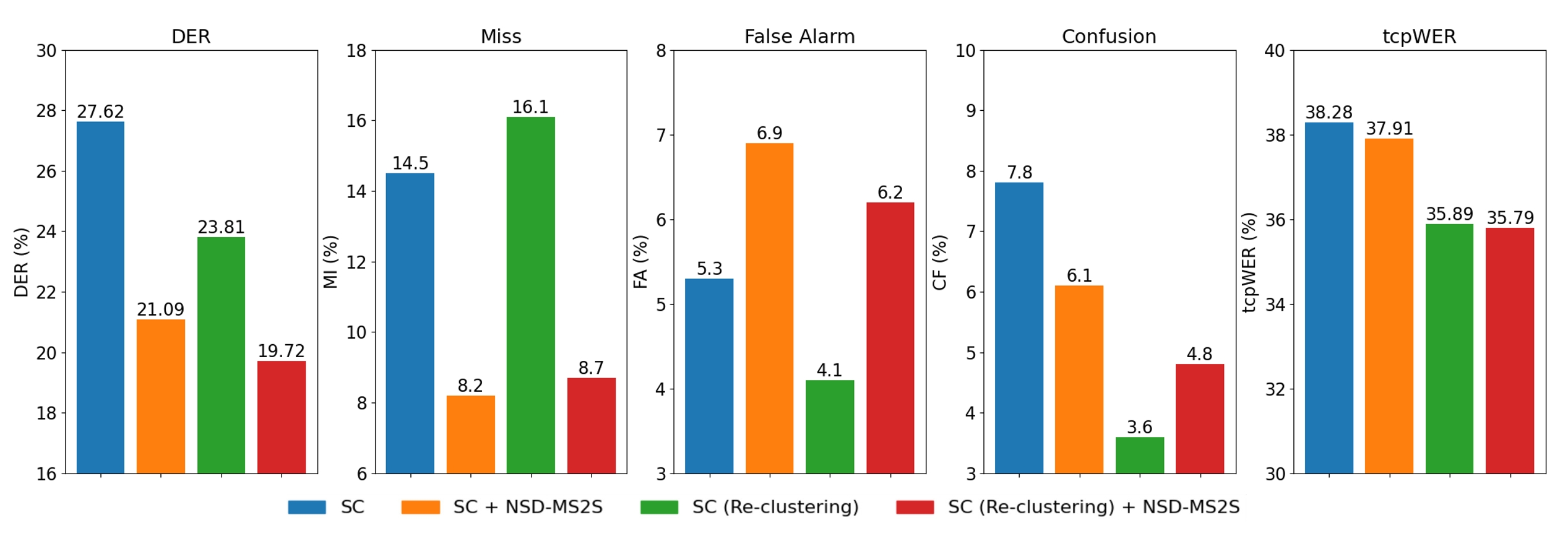}
	\vspace{-0.0cm}
	\caption{Detailed DERs ($\%$) and tcpWERs ($\%$) of different diarization systems on the single-channel NOTSOFAR-1 eval-small dataset. The tcpWER results are obtained from DCF-DS ($T_{\text{W}}$ = 3s, $N_{\text{W}}$ = 3) without using RTM.}
	\label{fig5}
	\vspace{-0.2cm}
\end{figure*}
\begin{table}[t]
	\renewcommand\arraystretch{1.25}
	\newcolumntype{L}[1]{>{\raggedright\arraybackslash}p{#1}}
	\newcolumntype{C}[1]{>{\centering\arraybackslash}p{#1}}
	\newcolumntype{R}[1]{>{\raggedleft\arraybackslash}p{#1}}
	\centering
	\caption{tcpWER ($\%$) of DCF-DS systems under different configurations on the single-channel NOTSOFAR-1 eval-small dataset. For the back-end speech recognition system, we uniformly use the Whisper ``large-v3'' model \cite{radford2023robust}. $T_{\text{W}}$ means the length of the sliding window.}
	\label{tab:2}\medskip
	\resizebox{9 cm}{!}{\begin{tabular}{l|c|c|c}
			\toprule[1 pt]
			\textbf{Front-end Systems}&\textbf{Training Datasets}&$T_{\text{W}}$&\textbf{tcpWER} ($\%$)\\
			\midrule
			NOTSOFAR-1 Baseline \cite{vinnikov24_interspeech} &\multirow{2}*{NOTSOFAR-Sim}&\multirow{2}*{3s}&41.4 \\
			DCF-DS&&&37.31 \\
			\midrule
			DCF-DS &\multirow{2}*{NOTSOFAR-Sim}&\multirow{2}*{12.8s}&34.93\\
			DCF-DS (i-vector)&&&37.35\\
			\midrule
			DCF-DS&\multirow{4}*{\makecell[c]{NOTSOFAR-Sim \\ + NF-Sim} }&\multirow{4}*{12.8s}&34.51 \\
			+ MIMO-SE Stage 2&&&33.81 \\
			+ MIMO-SE Stage 3 &&&33.97 \\
			+ Re-clustering&&&\textbf{31.72}\\
			\bottomrule[1 pt]	
	\end{tabular}}
	\vspace{-0.3 cm}
\end{table}

Table \ref{tab:2} presents the performance of DCF-DS systems under different configurations. First, comparing the first and second blocks, we illustrate the impact of window length with the same NOTSOFAR-1 simulated training dataset. When we extend the window length from the baseline configuration of 3 seconds to 12.8 seconds (800 frames with a 16-ms frame shift), the maximum number of speakers per window ($N_{\text{W}}$) increases to 4. The results show that increasing the window length effectively improves system performance, reducing the tcpWER from 37.31$\%$ to 34.93$\%$.
In the second block, we present results for a variant of the DCF-DS system where the time soft mask $ \hat{\mathbf{Y}}_{\text{W}}^{(i)} $ is replaced with the i-vectors. This configuration is similar to the TS-SEP approach \cite{boeddeker2024ts}, where speaker embeddings are used to associate outputs with specific speaker identities, facilitating target-speaker separation. As shown in the results, relying solely on speaker embeddings without any temporal boundary information leads to a performance decline (tcpWER increasing from 34.93$\%$ to 37.35$\%$). This highlights the importance of the speaker diarization component in DCF-DS, as removing it significantly impacts performance, and underscores the positive role of time boundaries in enhancing the separated speech quality.
Furthermore, \cite{boeddeker2024ts} reports that directly training TS-SEP with an 8-output node structure results in non-convergence. In contrast, the 4-output node structure used in DCF-DS (i-vector) converges successfully, indicating that the proposed DCF-DS framework effectively mitigates the SDCI issue.
In the third block, we first present results obtained by introducing additional training data simulated from the near-field (NF) close-talking recordings corpus, referred to as ``NF-Sim''. Adding more simulated data yields only a slight improvement (tcpWER from 34.93$\%$ to 34.51$\%$), maybe because the original 1000-hour simulated data already provide sufficient coverage, so additional data do not introduce significant new information. After passing the separated streams through the MIMO-SE module, we observe an enhancement in speech quality, which effectively improves ASR performance (tcpWER from 34.51$\%$ to 33.81$\%$). However, when the T-F masks predicted by the first MIMO-SE module are fed iteratively back into MIMO-SE for a multi-stage process, as shown in Fig. \ref{fig3}, there is no additional gain (tcpWER remains at 33.97$\%$). This indicates that a single pass through MIMO-SE achieves sufficient speech quality enhancement, with no further improvement from additional iterations.
Finally, we refine the initial speaker diarization priors $ \mathbf{S}_{\text{G}} $ through a re-clustering process (the details will be analyzed in the subsequent results). Initializing the DCF-DS system with these improved diarization priors leads to a substantial performance gain, reducing tcpWER from 33.81$\%$ to 31.72$\%$. Overall, our system achieves a relative improvement of 23.4$\%$ compared with the baseline, reducing tcpWER from 41.4$\%$ to 31.72$\%$.


Fig. \ref{fig5} presents the detailed DERs of different speaker diarization systems, along with the corresponding tcpWER results obtained from DCF-DS ($T_{\text{W}}$ = 3s, $N_{\text{W}}$ = 3) without using RTM. Comparing the DERs of ``SC'' and ``SC + NSD-MS2S'', we observe a significant improvement (from 27.62$\%$ to 21.09$\%$). This improvement mainly comes from reductions in MI (a decrease of 6.3$\%$, from 14.5$\%$ to 8.2$\%$). However, the tcpWER only shows a modest reduction of 0.37$\%$ (from 38.28$\%$ to 37.91$\%$). This suggests that while speaker diarization performance shows considerable improvement, it does not always lead to a significant enhancement in speech recognition performance, consistent with the conclusion in \cite{taherian2024multi}.
Moreover, comparing the ``SC (Re-clustering)'' and ``SC + NSD-MS2S'', we observe that although the diarization result after re-clustering (DER = 23.81$\%$) is worse than the first NSD-MS2S result (DER = 21.09$\%$), the corresponding tcpWER shows a significant improvement (35.89$\%$ versus 37.91$\%$). This improvement can primarily be attributed to the reduction in CF (from 6.1$\%$ to 3.6$\%$), which nearly yields an equivalent gain in speech recognition performance (an absolute decrease of 2.02$\%$). Therefore, CF is a particularly important metric in speaker diarization results, especially for multi-speaker speech recognition. 
Furthermore, comparing the ``SC (Re-clustering)'' and ``SC (Re-clustering) + NSD-MS2S'', we observe that although the overall DER of NSD-MS2S improves compared to the initialized SC system (decreasing from 23.81$\%$ to 19.72$\%$), its CF slightly increases (from 3.6$\%$ to 4.8$\%$), which is harmful to the final speech recognition performance. This may be due to the fact that NSD-MS2S, compared to SC, reduces much of the MI, leading to the recovery of more speakers and inevitably increasing CF. From the perspective of speech recognition, initializing with NSD-MS2S still results in a slight improvement (tcpWER decreases from 35.89$\%$ to 35.79$\%$), likely because the significant reduction in MI by NSD-MS2S compensates for the increase in CF.
Finally, comparing the ``SC'' and ``SC (Re-clustering)'', we find that the use of re-clustering substantially improves diarization performance (DER reduced from 27.62$\%$ to 23.81$\%$), with the main gain coming from the decrease in CF (from 7.8$\%$ to 3.6$\%$). This demonstrates that the DCF-DS framework effectively leverages separated speech to enhance diarization performance, thereby bringing a significant improvement in downstream speech recognition.
\begin{table}[t]
	\renewcommand\arraystretch{1.25}
	\newcolumntype{L}[1]{>{\raggedright\arraybackslash}p{#1}}
	\newcolumntype{C}[1]{>{\centering\arraybackslash}p{#1}}
	\newcolumntype{R}[1]{>{\raggedleft\arraybackslash}p{#1}}
	\centering
	\caption{tcpWER ($\%$) comparison of the single systems (excluding fusion systems) submitted by other teams on the NOTSOFAR-1 single-channel track evaluation set.}
	\label{tab:4}\medskip
	\resizebox{8 cm}{!}{\begin{tabular}{llc}
			\toprule[1 pt]
			\textbf{Front-end Systems}&\textbf{Back-end Systems}&\textbf{tcpWER} ($\%$)\\
			\midrule
			Baseline \cite{vinnikov24_interspeech} &Whisper large-v3&41.4 \\
			\midrule
			Oracle Segment \cite{vinnikov24_interspeech} &Whisper large-v3&39.7 \\
			\midrule
			EEND-TA \cite{broughton2024fano}&Whisper large-v3&44.2\\
			\midrule
			PixIT \cite{kalda2024totato, kalda24_odyssey} &Whisper large-v3 (FT) \cite{kalda2024totato}&41.2\\
			\midrule
			EEND (WavLM) \cite{polok24_chime}& \makecell[l]{Whisper large-v3 \\(FDDT Large+FT) \cite{polok24_chime}} & 40.1\\
			\midrule
			\makecell[l]{CSS (Conformer)  \\+ SC (NME) \cite{hirano2024naist}}  & \makecell[l]{Zipformer Transducer \\ (WavLM-large) \cite{hirano2024naist}} & 36.6\\
			\midrule
			\makecell[l]{CSS (TF-GridNet)  \\+ SC (ML-NME) \cite{hu24_chime}}  & \makecell[l]{Whisper large-v2 \\ (ASR Techniques) \cite{hu24_chime}} & 33.5\\
			\midrule
			\makecell[l]{CSS (WavLM)  \\+ SC (ResNet293) \cite{huang2024npu}}  & \makecell[l]{Whisper large-v2 (FT) \cite{huang2024npu}} & 34.6\\
			\midrule
			\multirow{2}*{\makecell[l]{DCF-DS}}  & Whisper large-v3& \textbf{31.7}\\
			& Enhanced Whisper \cite{niu24_chime}& \textbf{22.2}\\
			\bottomrule[1 pt]	
	\end{tabular}}
	\vspace{-0.3 cm}
\end{table}

Table \ref{tab:4} presents the results of single systems submitted by other teams in the single-channel track of the NOTSOFAR-1 challenge. It is important to note that since each system may have used different training datasets and incorporated various improvements for the back-end systems, we are not making a strict performance comparison between methods. Instead, our goal is to present the performance obtained by advanced methods in the field.
In \cite{broughton2024fano}, the system used the EEND with transformer-based attractor calculation (EEND-TA) \cite{samarakoon2023transformer} method for speaker diarization and employed Whisper ``large-v3'' as the back-end, achieving a tcpWER of 44.2$\%$. In \cite{kalda2024totato}, the system utilized the PixIT framework \cite{kalda24_odyssey} as the front-end and fine-tuned (FT) the Whisper ``large-v3'' on in-domain data for speech recognition, resulting in a tcpWER of 41.2$\%$. In \cite{polok24_chime}, the system used the EEND system combined with WavLM as the front-end, and applied the Frame-Level Diarization Dependent Transformations (FDDT) method for the back-end Whisper model, transforming it into a target-speaker ASR. The ASR model was also fine-tuned on in-domain data, achieving a tcpWER of 40.1$\%$. Systems \cite{hirano2024naist}, \cite{hu24_chime} and \cite{huang2024npu} all adopted the same ``CSS + SC'' approach as the baseline, with optimizations made to various modules. The system in \cite{hirano2024naist} adopted the Zipformer transducer as the ASR model with WavLM-large features, and used the Conformer CSS and normalized maximum eigengap-based (NME) spectral clustering, achieving a tcpWER of 36.6$\%$.
The system in  \cite{hu24_chime} utilized the TF-GridNet \cite{wang2023tf} and multi-level normalized maximum eigengap-based (ML-NME) spectral clustering methods, and incorporated some ASR techniques such as rescoring using a pre-trained language model (LM), achieving a tcpWER of 33.5$\%$. The system in \cite{huang2024npu} combined WavLM with the CSS component in the front-end, and enhanced the speaker embedding extraction model with the ResNet293 \cite{chen2022sjtu}. The system in \cite{huang2024npu} also fine-tuned the back-end model, ultimately achieving a tcpWER of 34.6$\%$.
The proposed DCF-DS system achieves the lowest tcpWER of 31.7$\%$ when using the baseline Whisper ``large-v3'' as the back-end ASR model. Furthermore, when the improved back-end model, Enhanced Whisper \cite{niu24_chime}, is employed, the performance is further enhanced, achieving a tcpWER of 22.2$\%$. Ultimately, our system won first place in the single-channel track of the NOTSOFAR-1 Challenge\footnote{https://www.chimechallenge.org/challenges/chime8/task2/results}.
\subsection{Performance Comparison on LibriCSS}
\label{subsection44: LB}
\begin{table}[t]
	\renewcommand\arraystretch{1.25}
	\newcolumntype{L}[1]{>{\raggedright\arraybackslash}p{#1}}
	\newcolumntype{C}[1]{>{\centering\arraybackslash}p{#1}}
	\newcolumntype{R}[1]{>{\raggedleft\arraybackslash}p{#1}}
	\centering
	\caption{The cpWERs ($\%$) of DCF-DS systems under different configurations on the single-channel LibriCSS dataset. $T_{\text{W}}$ means the length of the sliding window.}
	\label{tab:5}\medskip
	\resizebox{9 cm}{!}{\begin{tabular}{l|c|c|c|c}
			\toprule[1 pt]
			\textbf{Front-end Systems}&\textbf{Training Datasets}&$T_{\text{W}}$&\textbf{ASR}&\textbf{cpWER} ($\%$)\\
			\midrule
			Oracle Segment \cite{taherian2024multi}&-&-&E2E&19.72 \\
			\midrule
			\multirow{2}*{DCF-DS}&\multirow{2}*{NOTSOFAR-Sim}&3s&E2E&8.28 \\
			&&12.8s&E2E&7.62 \\
			\midrule
			DCF-DS&\multirow{5}*{\makecell[c]{NOTSOFAR-Sim \\ + NF-Sim}}&\multirow{5}*{12.8s}&E2E&7.53 \\
			DCF-DS&&&E2E-SSL&5.67\\
			+ MIMO-SE Stage 2&&&E2E-SSL&5.53\\
			+ MIMO-SE Stage 3&&&E2E-SSL&5.54\\
			+ Re-clustering&&&E2E-SSL&\textbf{4.43}\\
			\bottomrule[1 pt]	
	\end{tabular}}
	\vspace{-0.3 cm}
\end{table}
Table \ref{tab:5} presents the performance of the proposed methods on the single-channel LibriCSS dataset. It is important to note that we test the pre-trained model directly on the LibriCSS dataset without domain adaptation. This is relatively challenging since the LibriCSS dataset includes recordings with overlap ratios ranging from 0$\%$ to 40$\%$. It also contains utterances with long periods of silence (0L subset), which are almost unseen in our training set. The results show that DCF-DS still achieves strong performance on the LibriCSS dataset, with trends consistent with those observed in Table \ref{tab:2}. Specifically, longer window lengths, MIMO-SE module, and re-clustering all contribute to improvements, demonstrating the good generalization ability of the DCF-DS method. Finally, we achieve a cpWER of 4.43$\%$ on the single-channel LibriCSS dataset. From Tables \ref{tab:2} and \ref{tab:5}, it is shown that although the number of output nodes in DCF-DS is limited to 3 or 4, the method still performs well in multi-speaker scenarios (e.g., 4-8 speakers in NOTSOFAR-1 eval set and 8 speakers in LibriCSS dataset). This also indicates that DCF-DS effectively mitigates the SDCI issue.
\begin{table}[t]
	\renewcommand\arraystretch{1.25}
	\newcolumntype{L}[1]{>{\raggedright\arraybackslash}p{#1}}
	\newcolumntype{C}[1]{>{\centering\arraybackslash}p{#1}}
	\newcolumntype{R}[1]{>{\raggedleft\arraybackslash}p{#1}}
	\centering
	\caption{Literature comparison on single-channel LibriCSS dataset. Following \cite{boeddeker2024ts}, the ``Style" column outlines the system architecture. D, S, and A stand for diarization, separation, and ASR, respectively. The ``→" represents a cascade, and ``+" indicates coupling integration.}
	\label{tab:6}\medskip
	\resizebox{8 cm}{!}{\begin{tabular}{l|l|c}
			\toprule[1 pt]
			\textbf{Systems}&\textbf{Style}&\textbf{cpWER} ($\%$)\\
			\midrule
			AHC \cite{raj2021integration}&D$\rightarrow$A&36.7\\
			VBx \cite{raj2021integration}&D$\rightarrow$A&33.4\\
			SC \cite{raj2021integration}&D$\rightarrow$A&31.0\\
			Transcribe-to-Diarize \cite{kanda2022transcribe}&End-to-End&11.6\\
			ADEnet $\rightarrow$ E2E \cite{delcroix2021speaker}&D$\rightarrow$S$\rightarrow$A&18.8\\
			TS-SEP $\rightarrow$ Mask $\rightarrow$ E2E \cite{boeddeker2024ts}&D + S $\rightarrow$ A&11.61\\
			TS-SEP $\rightarrow$ Mask $\rightarrow$ E2E-SSL \cite{boeddeker2024ts}&D + S $\rightarrow$ A&7.81\\
			CSS-AD $\rightarrow$ Whisper\cite{von2024meeting}& S$\rightarrow$A$\rightarrow$D&7.2\\
			CSS-AD $\rightarrow$ E2E-SSL\cite{von2024meeting}& S$\rightarrow$A$\rightarrow$D&6.2\\
			\midrule
			DCF-DS $\rightarrow$ E2E&\multirow{2}*{D + S $\rightarrow$ A}&6.30\\
			DCF-DS  $\rightarrow$ E2E-SSL&&\textbf{4.43}\\
			\bottomrule[1 pt]	
	\end{tabular}}
	\vspace{-0.3 cm}
\end{table}

Table \ref{tab:6} presents the cpWER results of representative methods on the LibriCSS dataset under the single-channel constraint. The first three rows show the recognition results using different conventional clustering methods as the front-end, including agglomerative hierarchical clustering (AHC) \cite{AHC}, variational Bayesian hidden Markov model (BHMM) with x-vector sequences (VBx)  \cite{VBxdihard}, and spectral clustering (SC) \cite{park2019auto}. It can be observed that, without separation, the recognition performance is generally poor (cpWER over 30$\%$). The Transcribe-to-Diarize \cite{kanda2022transcribe} method uses an ASR model to generate timestamps, achieving a cpWER of 11.6$\%$. The system in \cite{delcroix2021speaker}, which combines TS-VAD and speech extraction, achieves a cpWER of 18.8$\%$. The TS-SEP method reaches cpWERs of 11.61$\%$ and 7.81$\%$ using E2E and E2E-SSL as ASR models, respectively. CSS-AD \cite{von2024meeting} achieves a cpWER of 6.2$\%$ based on the E2E-SSL ASR model. Finally, our proposed DCF-DS achieves cpWER of 4.43$\%$ with the E2E-SSL ASR model, demonstrating the state-of-the-art recognition performance on the single-channel LibriCSS dataset.

\section{Conclusion}
\label{section5: Conclusion}
In this paper, we propose the DCF-DS framework, which addresses the SDCI problem by sequential joint training of speaker diarization and speech separation. Additionally, we introduce the MIMO-SE module and re-clustering approach based on separation results, further enhancing back-end recognition performance. Our final method achieves state-of-the-art results on both the single-channel NOTSOFAR-1 and LibriCSS datasets. In future work, we aim to explore the application of the DCF-DS framework in multi-channel scenarios.

\bibliographystyle{IEEEtran}
\bibliography{mybib}
\end{document}